\documentclass[aps,prl,twocolumn,showpacs,amsmath,amssymb]{revtex4-1}
\usepackage{amsmath}
\usepackage{graphicx}
\usepackage{subfigure}
\usepackage{epstopdf}
\usepackage{color}
\usepackage{multirow}
\usepackage{setspace}
\usepackage{overpic}
\usepackage{amssymb}
\usepackage[dvipdfmx, bookmarksnumbered, pdfstartview=FitH,colorlinks,urlcolor=blue, citecolor=blue,linkcolor=blue] {hyperref}
\usepackage{lineno}
\usepackage{bm}
\usepackage{rotating}
\usepackage[utf8]{inputenc}
\let\oldequation\equation
\let\oldendequation\endequation
\renewenvironment{equation}
  {\linenomathNonumbers\oldequation}
  {\oldendequation\endlinenomath}

\begin{document}

\title{\bf \boldmath
Measurements of Absolute Branching Fractions of Fourteen Exclusive Hadronic $D$ Decays to $\eta$
}

\author{
M.~Ablikim$^{1}$, M.~N.~Achasov$^{10,c}$, P.~Adlarson$^{64}$, S. ~Ahmed$^{15}$, M.~Albrecht$^{4}$, A.~Amoroso$^{63A,63C}$, Q.~An$^{60,48}$, ~Anita$^{21}$, X.~H.~Bai$^{54}$, Y.~Bai$^{47}$
% ...
\\
\vspace{0.2cm}
(BESIII Collaboration)\\
\vspace{0.2cm} {\it
$^{1}$ Institute of High Energy Physics, Beijing 100049, People's Republic of China\\
% ...
$^{4}$ Bochum Ruhr-University, D-44780 Bochum, Germany\\
% ...
$^{10}$ G.I. Budker Institute of Nuclear Physics SB RAS (BINP), Novosibirsk 630090, Russia\\
% ...
$^{15}$ Helmholtz Institute Mainz, Johann-Joachim-Becher-Weg 45, D-55099 Mainz, Germany\\
% ...
$^{21}$ Indian Institute of Technology Madras, Chennai 600036, India\\
% ...
$^{47}$ Southeast University, Nanjing 211100, People's Republic of China\\
$^{48}$ State Key Laboratory of Particle Detection and Electronics, Beijing 100049, Hefei 230026, People's Republic of China\\
% ...
$^{54}$ University of Jinan, Jinan 250022, People's Republic of China\\
% ...
$^{60}$ University of Science and Technology of China, Hefei 230026, People's Republic of China\\
% ...
$^{63}$ (A)University of Turin, I-10125, Turin, Italy; (B)University of Eastern Piedmont, I-15121, Alessandria, Italy; (C)INFN, I-10125, Turin, Italy\\
$^{64}$ Uppsala University, Box 516, SE-75120 Uppsala, Sweden\\
% ...
$^{c}$ Also at the Novosibirsk State University, Novosibirsk, 630090, Russia\\
% ...
}
}

\begin{abstract}
Using $2.93\,\rm fb^{-1}$ of $e^+e^-$ collision data taken
at a center-of-mass energy of 3.773\,GeV with the BESIII detector,
we report the first measurements of the absolute branching fractions of fourteen
hadronic $D^{0(+)}$ decays to exclusive final states with an $\eta$, e.g., $D^0\to K^-\pi^+\eta$, $K^0_S\pi^0\eta$, $K^+K^-\eta$, $K^0_SK^0_S\eta$, $K^-\pi^+\pi^0\eta$, $K^0_S\pi^+\pi^-\eta$, $K^0_S\pi^0\pi^0\eta$, and $\pi^+\pi^-\pi^0\eta$; $D^+\to K^0_S\pi^+\eta$, $K^0_SK^+\eta$, $K^-\pi^+\pi^+\eta$, $K^0_S\pi^+\pi^0\eta$, $\pi^+\pi^+\pi^-\eta$, and $\pi^+\pi^0\pi^0\eta$. Among these decays, the $D^0\to K^-\pi^+\eta$ and $D^+\to K^0_S\pi^+\eta$ decays have the largest branching fractions, which are
$\mathcal{B} (D^0\to K^-\pi^+\eta )=(1.853\pm0.025_{\rm stat}\pm0.031_{\rm syst})\%$ and
$\mathcal{B}(D^+\to K^0_S\pi^+\eta)=(1.309\pm0.037_{\rm stat}\pm0.031_{\rm syst})\%$, respectively.
The $CP$ asymmetries for the six decays with highest event yields are determined, and no statistically significant $CP$ violation is found.
\end{abstract}


\maketitle

\oddsidemargin  -0.2cm
\evensidemargin -0.2cm

Hadronic $D$ decays provide an ideal platform to explore strong and weak effects in decays of hadrons with charm or bottom quarks.
Tests of lepton flavor universality (LFU) with semileptonic $B$ decays are important to explore new physics beyond the SM.
In recent years,
the branching fraction (BF) ratios
${\mathcal R}_{\tau/\ell}={\mathcal B}_{B\to \bar D^{(*)}\tau^+{\bar \nu}_\tau}/{\mathcal B}_{B\to \bar D^{(*)}\ell^+{\bar \nu}_\ell}$~($\ell=\mu$, $e$) measured by BaBar, Belle, and LHCb~\cite{babar_1,babar_2,lhcb_1,belle2015,belle2016,lhcb_1a,belle2019,belle2019_2} were found to deviate from the standard model (SM)
prediction by $3.1\sigma$~\cite{hflav2018}.
It is argued in Ref.~\cite{lhcbnote} that the exclusive hadronic $D^{0(+)}$ decays to $\eta$ are key potential backgrounds in these tests.
However, the known exclusive $D^0$ and $D^+$ decays to final states
with an $\eta$ meson only account for 44\% and 16\% of their corresponding inclusive rates~\cite{pdg2018}, respectively.
In particular, the BFs for the decays $D\to\bar K\pi\eta$, $K\bar K\eta$, $\bar K\pi\pi\eta$, and $\pi\pi\pi\eta$ (excluding narrow peaks $K^0_S$, $\eta$, $\omega$, $\eta^{\prime}$, and $\phi$ in individual mass spectra) are poorly known,
except for relative measurements of $D^0\to K^-\pi^+\eta$~\cite{belle-kpieta} and $D^0\to K^0_S\pi^0\eta$~\cite{cleo-kspi0eta}.
Measurements of the BFs of these decays are crucial to address the tensions found in LFU tests with semileptonic $B$ decays.
Furthermore, combining the measured BFs
with the corresponding amplitude analysis results gives important information on two-body hadronic $D$ decays. This is essential for improving the understanding of
quark U-spin~\cite{Kingsley,Gronau,zzxing} and SU(3)-flavor symmetry breaking effects, thereby benefiting theoretical predictions of $D^0\bar D^0$ mixing and charge-parity ($CP$) violation in $D$ decays~\cite{ref5,theory_a,theory_5,theory_4,theory_3,theory_2,theory_1,zzxing}.

Studies of $CP$ violation in the weak decays of hadrons are powerful tools for understanding physics within the SM and searches for physics beyond it.  The $CP$ violation in $D$ decays is predicted to be up to a few times $10^{-3}$~\cite{ref1,ref2,ref3,ref4,ref5,ref6,ref7}
and has been recently observed to be $(1.54\pm0.29)\times 10^{-3}$ in $D^0\to K^+K^-$ and $\pi^+\pi^-$ decays by LHCb~\cite{lhcb_D_CP}.
However, knowledge of $CP$ violation in $D$ decays is still very limited.
Searching for $CP$ asymmetries in hadronic $D$ decays, which have been much less explored than (semi-)leptonic decays, allows for a more comprehensive understanding of $CP$ violation in the $D$ sector.

This Letter reports the first measurements of the absolute BFs for the decays
$D^0\to K^-\pi^+\eta$, $K^0_S\pi^0\eta$, $K^+K^-\eta$, $K^0_SK^0_S\eta$, $K^-\pi^+\pi^0\eta$, $K^0_S\pi^+\pi^-\eta$, $K^0_S\pi^0\pi^0\eta$, $\pi^+\pi^-\pi^0\eta$, and $D^+\to K^0_S\pi^+\eta$, $K^0_SK^+\eta$, $K^-\pi^+\pi^+\eta$, $K^0_S\pi^+\pi^0\eta$, $\pi^+\pi^+\pi^-\eta$, $\pi^+\pi^0\pi^0\eta$. Throughout this Letter, the charge conjugate processes are implied unless stated otherwise.
In addition, the $CP$ asymmetries are determined for the six decays with the highest yields.
To avoid double-counting previously measured decays,
the narrow peaks for the $K^0_S$, $\eta$, $\omega$, $\eta^{\prime}$, and $\phi$
are removed from the mass spectra of the $\pi^{+(0)}\pi^{-(0)}$, $\pi^+\pi^-\pi^0$, $\pi^+\pi^-\pi^0$, $\pi^{+(0)}\pi^{-(0)}\eta$, and $K^+K^-$\,(or\,$\pi^+\pi^-\pi^0$) combinations,
respectively.

The data sample was collected with the BESIII detector at a center-of-mass energy
$\sqrt s=$ 3.773~GeV and has an integrated luminosity of 2.93\,fb$^{-1}$~\cite{lum_bes3}.
Details about the design and performance of the BESIII detector are given in Ref.~\cite{BESIII}.
The Monte Carlo (MC) simulated events are produced with
a {\sc geant4}-based~\cite{geant4} detector simulation software package. An inclusive MC sample, including  $D^0\bar{D}^0$, $D^+D^-$, and non-$D\bar{D}$ decays of the $\psi(3770)$, initial state radiation~(ISR) production of the $\psi(3686)$ and $J/\psi$, and the processes
$e^+e^-\to q\bar{q}$~($q=u,d,s$) and $e^+e^-\to(\gamma)\ell^+\ell^-$ ($\ell=e,\mu,\tau$), is produced to determine the detection efficiencies and to estimate any potential backgrounds. The production of the charmonium states is simulated by the MC generator {\sc kkmc}~\cite{kkmc}.  The measured decay modes of the charmonium states are generated using {\sc evtgen}~\cite{evtgen} with BFs from the Particle Data Group~\cite{pdg2018}, and the remaining unknown decay modes are generated by {\sc lundcharm}~\cite{lundcharm}.

The BFs of the hadronic $D$ ($D^0$ or $D^+$) decays are measured via the reaction chain $e^+e^-\to \psi(3770)\to D\bar D$.
If a $\bar D$ meson is fully reconstructed, it is called a single-tag (ST) $\bar D$ meson.
The ST $D^-$ mesons are reconstructed via the decays
$D^-\to K^{+}\pi^{-}\pi^{-}$,
$K^0_{S}\pi^{-}$, $K^{+}\pi^{-}\pi^{-}\pi^{0}$, $K^0_{S}\pi^{-}\pi^{0}$, $K^0_{S}\pi^{+}\pi^{-}\pi^{-}$,
and $K^{+}K^{-}\pi^{-}$, while
the ST $\bar D^0$ mesons are reconstructed using the
decays $\bar D^0\to K^+\pi^-$, $K^+\pi^-\pi^0$, and $K^+\pi^-\pi^-\pi^+$.
If a signal decay is fully reconstructed
in the system recoiling against an ST $\bar D$ meson,
the candidate event is called a double-tag (DT) event.
The BF of the signal decay is given by
\begin{equation}
\label{eq:br}
{\mathcal B}_{{\rm sig}} = {N_{\rm DT}}/({N_{\rm ST}}\cdot\epsilon_{{\rm sig}}),
\end{equation}
where
${N_{\rm ST}}=\sum_i N_{{\rm ST}}^i$ and ${ N_{\rm DT}}$
are the total ST and DT yields in data, respectively, and
$\epsilon_{{\rm sig}} = \sum_i (N^i_{{\rm ST}}\cdot\epsilon^i_{{\rm DT}}/\epsilon^i_{{\rm ST}})/{ N_{\rm ST}}$ is the effective efficiency for detecting the signal decay, averaged over tag mode $i$, where
$\epsilon_{{\rm ST}}$ and $\epsilon_{{\rm DT}}$ are the efficiencies for detecting ST and DT candidates,
respectively.

We use the same selection criteria for $K^\pm$, $\pi^\pm$, $K^0_S$, $\gamma$, and $\pi^0$ as were used in Refs.~\cite{epjc76,cpc40,bes3-pimuv,bes3-Dp-K1ev,bes3-etaetapi,bes3-omegamuv,bes3-etamuv}.
Candidates for $\eta$ are reconstructed from $\gamma\gamma$ pairs with invariant mass within
$(0.515,0.570)$\,GeV/$c^2$. To improve resolution, a one-constraint kinematic fit is imposed on
each $\gamma\gamma$ pair to constrain their invariant mass at $\eta$ nominal mass~\cite{pdg2018}.
For $\bar D^0\to K^+\pi^-\pi^-\pi^+$ tags, the $\bar D^0\to
K^0_SK^\pm\pi^\mp$ decays are rejected if the mass of any $\pi^+\pi^-$
pair falls in the range $(0.478,0.518)$ GeV/$c^2$.

Tagging $\bar D$ (signal $D$) mesons are identified by two variables, the energy difference $\Delta E_{\rm tag\,(sig)} \equiv E_{\rm tag\,(sig)} - E_{\rm b}$ and the beam-constrained mass $M_{\rm BC}^{\rm tag\,(sig)} \equiv \sqrt{E^{2}_{\rm b}-|\vec{p}_{\rm tag\,(sig)}|^{2}}$, where
tag\,(sig) represents the tagging $\bar D$ (signal $D$),
$E_{\rm b}$ the beam energy, and $\vec{p}_{\rm tag\,(sig)}$ and $E_{\rm tag\,(sig)}$ the momentum and energy of the $\bar D\,(D)$ candidate in the $e^+e^-$ rest frame. For each tag (signal) mode, if there are multiple combinations, only the one with the minimum $|\Delta E_{\rm tag\,(sig)}|$ is kept for further analysis.
The $\bar D$ tags are required to satisfy
$\Delta E_{\rm tag}\in(-55,\, 40)$\,MeV for the modes containing $\pi^0$ in the final states
and $\Delta E_{\rm tag}\in(-25,\, 25)$\,MeV for the other modes.
The yields of ST $\bar D$ mesons are obtained from binned maximum likelihood fits to the $M_{\rm BC}^{\rm tag}$
distributions of the accepted ST candidates following Refs.~\cite{epjc76,cpc40,bes3-pimuv,bes3-Dp-K1ev,bes3-etaetapi}.
The total ST $D^-$ yield is $ N_{{\rm ST}\,D^-} = 1558159\pm2113_{\rm stat}$.
The total ST $\bar D^0$ yield is $ N_{{\rm ST}\,\bar D^0} = 2327839\pm1860_{\rm stat}$ for self-conjugate signal $D^0$ decays.
For the flavor specific signal decays $D^0\to K^-\pi^+\eta$ and $K^-\pi^+\pi^0\eta$, we remove doubly Cabibbo suppressed
decays from the ST selection, giving $ N_{{\rm ST}\,\bar D^0}=
2321430\pm1860_{\rm stat}$ for these decays.

For the signal $D$ decays recoiling against the $\bar D$ tags, tracks are selected
from the residual tracks that have not been used for the tag reconstruction.
The signal $D$ decays are selected by using the $\Delta E_{\rm sig}$ requirements as listed in Table~\ref{tab:DT}.
For the $\pi^{+}\pi^{-}[\pi^{0}\pi^{0}]$, $\pi^+\pi^-\pi^0$, $\pi^+\pi^-\pi^0$, $\pi^{+(0)}\pi^{-(0)}\eta$, and $K^+K^-$ combinations,
the $K^0_S$, $\eta$, $\omega$, $\eta^\prime$, and $\phi$ contributions are rejected
by requiring their invariant masses to be outside
$(0.468,0.528)\,{\rm GeV}/c^2$ $[(0.438,0.538)\,{\rm GeV}/c^2$],
$(0.498,0.578)\,{\rm GeV}/c^2$,
$(0.732,0.832)\,{\rm GeV}/c^2$,
$(0.908,1.008)\,{\rm GeV}/c^2$, and
$(0.990,1.390)\,{\rm GeV}/c^2$, respectively. These correspond to at least five times the fitted mass resolution away from individual nominal mass.
For $D^0\to K^0_S\pi^{+(0)}\pi^{-(0)}\eta$ [$\pi^+\pi^-\pi^0\eta$] decays,
no aforementioned mass requirements of the $K^0_S\,[\phi]$ are imposed on the $\pi^{+(0)}\pi^{-(0)}$~[$\pi^+\pi^-\pi^0$] combinations,
due to the small BFs and the limited phase space (PHSP) of the background channels $D^0\to K^0_SK^0_S\eta$ [$\phi\eta$].
The opening angle between signal $D$ and tagging $\bar D$ is required to be greater than $160^\circ$,
with a loss of (2-6)\% of the signal, to suppress mis-formed $D\bar D$ candidates.
For $D\to \bar K\pi\pi^0\eta$, the peaking backgrounds (PBKG) of
$D\to \bar K\pi\pi^0\pi^0$ are rejected if
any $\bar K\pi\pi^0\pi^0$ combinations satisfying
$\Delta E_{\bar K\pi\pi^0\pi^0}\in (-0.05,0.05)$ GeV and $M^{\bar K\pi\pi^0\pi^0}_{\rm BC}\in (1.83,1.89)$ GeV/$c^2$ can be found in the same candidate events.
The combination of these requirements rejects more than 75\% of the background and keeps (93-97)\% of the signal.

To determine the DT yields in the data ($N^{\rm fit}_{\rm DT}$), a two-dimensional (2D) unbinned maximum likelihood fit
is performed on the $M_{\rm BC}^{\rm tag}$ vs. $M_{\rm BC}^{\rm sig}$
distribution of the accepted DT candidates (See Fig. 1 of the
supplemental material~\cite{supplemental} for an example).
Signal events concentrate around $M_{\rm BC}^{\rm tag} = M_{\rm BC}^{\rm sig} = M_{D}$,
where $M_{D}$ is the nominal $D$ mass~\cite{pdg2018}.
Background events are divided into three categories.
The first one (named BKGI) is from events with correctly reconstructed $D$ ($\bar D$) and incorrectly
reconstructed $\bar D$ ($D$). They are spread along the lines around
$M_{\rm BC}^{\rm tag}$ or $M_{\rm BC}^{\rm sig} = M_{D}$.
The second one (named BKGII) is from events smeared along the diagonal,
which are mainly from the $e^+e^- \to q\bar q$ processes.
The third one (named BKGIII) comes from events with uncorrelated and
incorrectly reconstructed $D$ and $\bar D$.

In the 2D fit, the probability density functions (PDFs) of the backgrounds are constructed as
\begin{itemize}
\item
BKGI: $b(x)\cdot c_y(y;E_{\rm b},\xi_{y},\frac{1}{2}) + b(y)\cdot c_x(x;E_{\rm b},\xi_{x},\frac{1}{2})$,
\item
BKGII: $c_z(z;\sqrt{2}E_{\rm b},\xi_{z},\frac{1}{2}) \cdot g(k;0,\sigma_k)$, and
\item
BKGIII: $c_x(x;E_{\rm b},\xi_{x},\frac{1}{2}) \cdot c_y(y;E_{\rm b},\xi_{y},\frac{1}{2})$.
\end{itemize}
Here, $x=M_{\rm BC}^{\rm sig}$, $y=M_{\rm BC}^{\rm tag}$, $z=(x+y)/\sqrt{2}$, and $k=(x-y)/\sqrt{2}$.
The PDFs for signal, $a(x,y)$, $b(x)$, and $b(y)$, are described by the corresponding MC-simulated shapes.
$c_f(f;E_{\rm b},\xi_f,\frac{1}{2})$ is an ARGUS function~\cite{ARGUS} defined as
$A_f\cdot f\cdot (1 - \frac {f^2}{E_{\rm b}^2})^{\frac{1}{2}} \cdot e^{\xi_f (1-\frac {f^2}{E_{\rm b}^2})}$,
where $f$ denotes $x$, $y$, or $z$; $E_{\rm b}$ is fixed at 1.8865 GeV/$c^2$; $A_f$ is a normalization factor;
and $\xi_f$ is a fit parameter.
$g(k;0,\sigma_k)$ is a Gaussian function with mean of zero and standard deviation $\sigma_k=\sigma_0 \cdot(\sqrt{2}E_{\rm b}-z)^p$,
where $\sigma_0$ and $p$ are two free parameters.
In addition to these backgrounds, for the decays $D^0\to K^0_S\pi^{+(0)}\pi^{-(0)}\eta$, $\pi^+\pi^-\pi^0\eta$, $K^-\pi^+\pi^0\eta$, $K^0_S\pi^0\pi^0\eta$,
and $D^+\to K^0_S\pi^+\pi^0\eta$, the yields and shapes of the PBKG components are fixed
based on MC simulations.
All other parameters are left free.

Combinatorial $\pi^+\pi^-$ pairs can also satisfy the $K^0_S$ selection criteria
and form peaking backgrounds around the $D$ mass in the $M_{\rm BC}^{\rm sig}$ distribution.
This kind of peaking background is estimated by the data events in the $K^0_S$ sideband region.
For $D^0\to K^0_S\pi^0\eta$, $K^0_S\pi^+\pi^-\eta$, $K^0_S\pi^0\pi^0\eta$, and
$D^+\to K^0_S\pi^+\eta$, $K^0_SK^+\eta$, $K^0_S\pi^+\pi^0\eta$,
one-dimensional (1D) signal and sideband regions are defined as
$M_{\pi^+\pi^-}\in(0.486,0.510)~{\rm GeV}/c^2$ and $M_{\pi^+\pi^-}\in (0.454,0.478)\cup(0.518,0.542)~{\rm GeV}/c^2$, respectively.
For $D^0\to K^0_SK^0_S\eta$,
2D signal and sideband regions are defined.
The 2D sideband 1~(2) regions are defined as the boxes
in which one~(two) of the two $\pi^+\pi^-$ combinations lie in the $K^0_S$ sideband regions
and the rest are located in the $K^0_S$ signal regions.
See Fig. 2 of the supplemental material~\cite{supplemental} as an example.

For the decays involving $K^0_S$, the net DT yields are obtained by
${N_{\rm DT}}
=N_{\rm DT}^{\rm fit} - \frac{1}{2}(N^{\rm fit}_{\rm sid{\text -}1}-N^{\rm fit}_{\rm sid{\text -}2})- \frac{1}{4}N^{\rm fit}_{\rm sid{\text -}2}
=N_{\rm DT}^{\rm fit} - \frac{1}{2}N^{\rm fit}_{\rm sid{\text -}1}+ \frac{1}{4}N^{\rm fit}_{\rm sid{\text -}2}$,
where $N_{\rm DT}^{\rm fit}$ and $N^{\rm fit}_{{\rm  sid}{\text -}i}$
are the fitted DT yields in the $K^0_S$ signal region and sideband $i$ region, respectively.
This relation has been verified based on MC simulation.
For the other decays, the net DT yields are $N^{\rm fit}_{\rm DT}$.

For each signal decay mode, the statistical significance is calculated by $\sqrt{-2{\rm ln ({\mathcal L_0}/{\mathcal L_{\rm max}}})}$, where ${\mathcal L}_{\rm max}$ and ${\mathcal L}_0$ are the maximum likelihoods with and without the signal component in the fits, respectively.
The effect of combinatorial $\pi^+\pi^-$ backgrounds in the $K^0_S$ signal regions has been considered for the decays involving $K^0_S$.
The statistical significances of the four decays with lowest yields, $D^0\to K^+K^-\eta$, $K^0_SK^0_S\eta$, $D^+\to K^0_SK^+\eta$, and $K^0_S\pi^+\pi^0\eta$, are $5.5\sigma$, $2.8\sigma$, $5.7\sigma$, and $8.4\sigma$, respectively; while those for the other decays are all greater than $10\sigma$.

To determine the signal efficiencies ($\epsilon^{}_{{\rm sig}}$), the $D\to \bar K\pi\eta$ decays are simulated with a modified data-driven generator BODY3~\cite{evtgen}, which was developed to simulate different intermediate states in data for a given three-body final state.
The Dalitz plot of $M^2_{\bar K\pi}$ vs. $M^2_{\pi\eta}$ found in data,
corrected for backgrounds and efficiencies, is taken as input for the BODY3 generator.
The efficiencies across the kinematic space are obtained with MC samples generated with the PHSP generator.
Intermediate states in the $D^0\to K^+K^-\eta$, $K^0_SK^0_S\eta$,
$K^0_S\pi^0\pi^0\eta$, and $D^+\to K^0_SK^+\eta$,
$K^0_S\pi^+\pi^0\eta$ decays cannot be determined due to limited
statistics; these decays are therefore simulated with the PHSP generator.
Each of the other decays is simulated with a mixed signal MC sample. Here, the decays generated with PHSP generator and the decays containing $K^*(892)$, $\rho(770)$, and $a_0(980)$ intermediate states are mixed with fractions obtained by examining the corresponding invariant mass spectra.
The data distributions for momenta and $\cos\theta$ (where $\theta$ is the
polar angle in the $e^+e^-$ rest frame) of the daughter particles, and the invariant masses of each of the two- and three-body particle combinations, agree with the MC simulations. The differences between the DT efficiencies obtained with the
BODY3 and PHSP generators will be assigned as a systematic uncertainty.

The values for ${ N_{{\rm DT}}}$, $\epsilon^{}_{{\rm sig}}$,
and the BFs of the signal decays are summarized in Table~\ref{tab:DT}.
The BF upper limit for $D^0\to K^0_SK^0_S\eta$ at 90\% confidence
level is determined to be $<2.4\times 10^{-4}$,
using the Bayesian approach after incorporating the systematic uncertainty~\cite{Stenson}.

\begin{figure*}[htp]
  \centering
\includegraphics[width=0.49\linewidth]{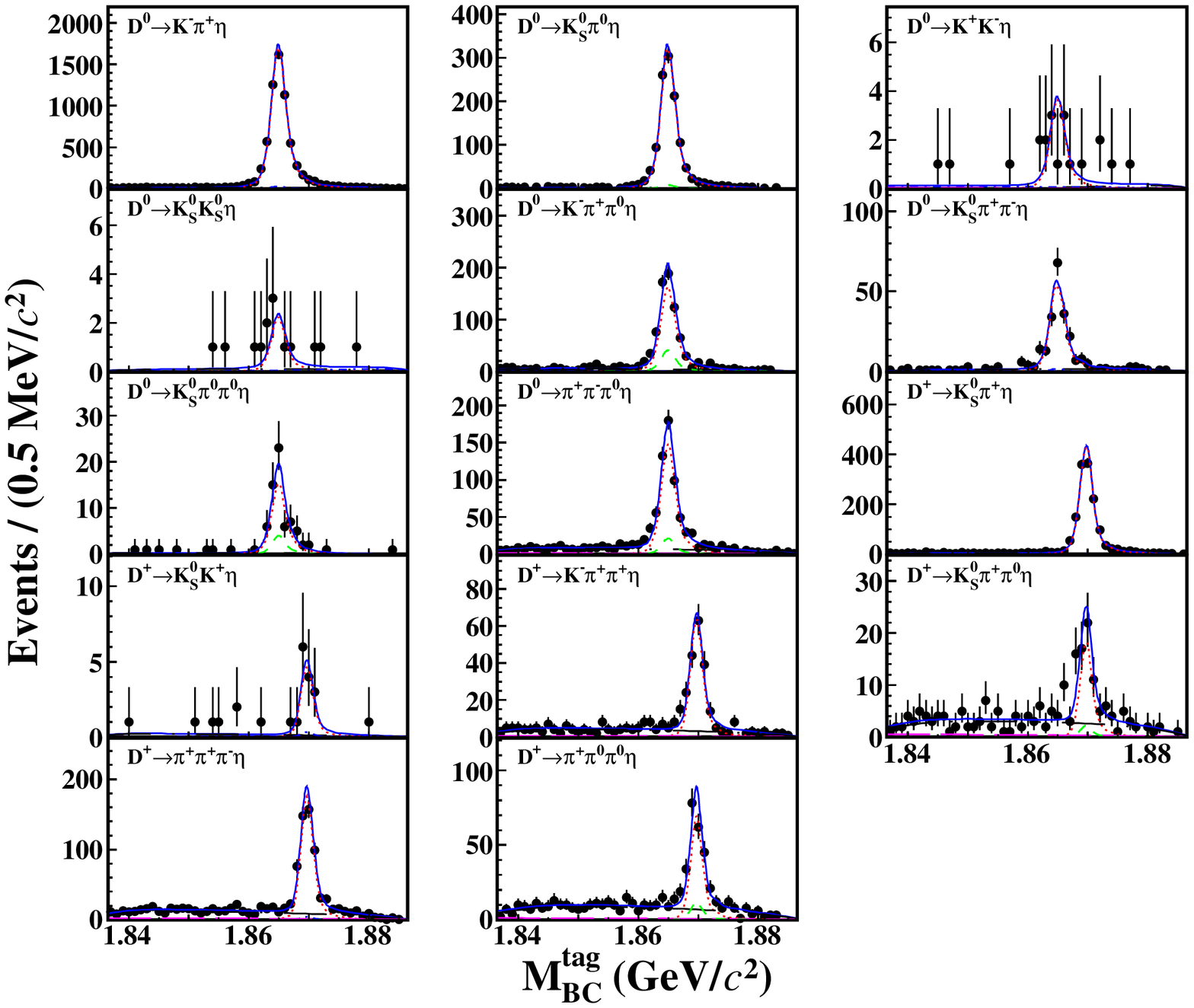}
\includegraphics[width=0.49\linewidth]{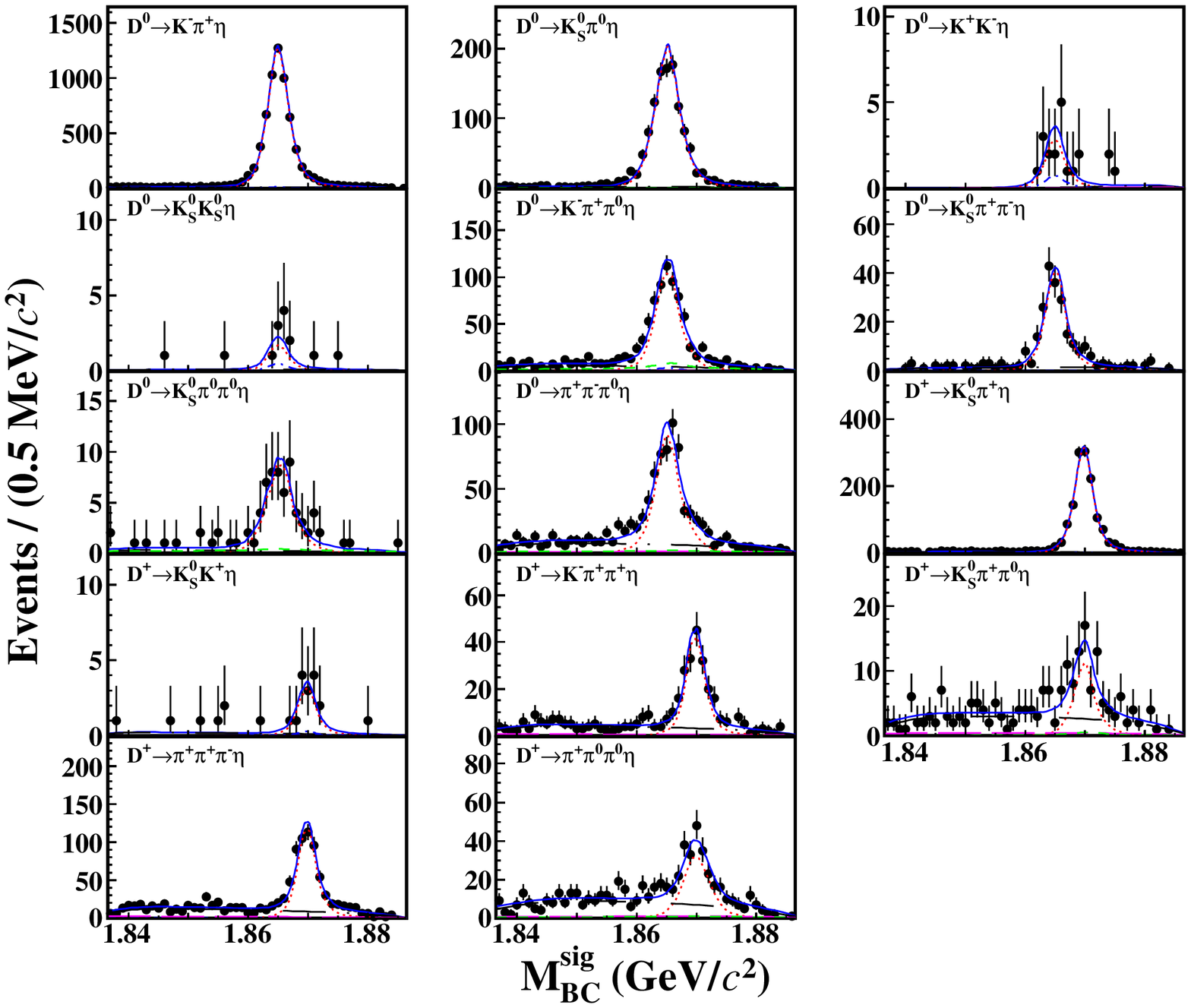}
  \caption{\small
Projections on $M^{\rm tag}_{\rm BC}$ and
$M^{\rm sig}_{\rm BC}$ of the 2D fits to the DT candidate events.
Data are shown as dots with error bars.
Blue solid, red-dotted, blue dot-dashed, black dot-long-dashed,
green long-dashed, and pink dashed curves denote the overall fit results,
signal, BKGI, BKGII, BKGIII, and PBKG components (see text),
respectively.
}
\label{fig:2Dfit}
\end{figure*}

\begin{table}[htp]
\centering
\caption{\small
Requirements on $\Delta E_{\rm sig}$,
net DT yields in data (${N_{{\rm DT}}}$),
detection efficiencies ($\epsilon_{\rm sig}$, including the BFs of $K^0_S$, $\eta$, and $\pi^0$ as well as correction factors described later),
and the obtained BFs (${\mathcal B}_{\rm sig}$).
Numbers in the first and second brackets are last two effective digits
of statistical and systematic uncertainties, respectively, for
${\mathcal B}_{\rm sig}$. The uncertainty is statistical only for ${ N_{{\rm DT}}}$.
The efficiency of $D^0\to K^+K^-\eta$ is significantly lower than that of $D^0 \to K^- \pi^+ \eta$,
because of lower selection efficiencies of $K^-$, $K^+$, and $\eta$ due to smaller PHSP as well as $\phi$ veto in $K^+K^-$ mass spectrum.
}\label{tab:DT}
\resizebox{!}{3.6cm}{
\begin{tabular}{lcrrr}
  \hline\hline
\multicolumn{1}{c} {Decay} &$\Delta E_{\rm sig}$& \multicolumn{1}{c} {${ N_{{\rm DT}}}$} & \multicolumn{1}{c} {$\epsilon_{\rm sig}$} &
  \multicolumn{1}{c} {${\mathcal B}_{\rm sig}$} \\
\multicolumn{1}{c} { } & (MeV) & \multicolumn{1}{c} {} & \multicolumn{1}{c} {(\%)} &
  \multicolumn{1}{c} {($\times 10^{-4}$)} \\  \hline
$D^0\to K^-\pi^+\eta$       &($-$37,\,36)&$6116.2\pm  81.8$ &14.22&185.3(25)(31)\\
$D^0\to K^0_S\pi^0\eta$     &($-$57,\,45)&$1092.7\pm  35.2$ & 4.66&100.6(34)(30)\\
$D^0\to K^+K^-\eta$         &($-$27,\,27)&$13.1  \pm~\,4.0$ & 9.53& 0.59(18)(05)\\
$D^0\to K^0_SK^0_S\eta$     &($-$29,\,28)&$7.3   \pm~\,3.2$ & 2.36& 1.33(59)(18)\\
$D^0\to K^-\pi^+\pi^0\eta$  &($-$44,\,36)&$576.5 \pm  28.8$ & 5.53& 44.9(22)(15)\\
$D^0\to K^0_S\pi^+\pi^-\eta$&($-$33,\,32)&$248.2 \pm  18.0$ & 3.80& 28.0(19)(10)\\
$D^0\to K^0_S\pi^0\pi^0\eta$&($-$56,\,41)&$64.7  \pm~\,9.2$ & 1.58& 17.6(23)(13)\\
$D^0\to \pi^+\pi^-\pi^0\eta$&($-$57,\,45)&$508.6 \pm  26.0$ & 6.76& 32.3(17)(14)\\
$D^+\to K^0_S\pi^+\eta$     &($-$36,\,36)&$1328.2\pm  37.8$ & 6.51&130.9(37)(31)\\
$D^+\to K^0_SK^+\eta$       &($-$27,\,27)&$13.6  \pm~\,3.9$ & 4.72& 1.85(52)(08)\\
$D^+\to K^-\pi^+\pi^+\eta$  &($-$33,\,33)&$188.0 \pm  15.3$ & 8.94& 13.5(11)(04)\\
$D^+\to K^0_S\pi^+\pi^0\eta$&($-$49,\,41)&$48.7  \pm~\,9.7$ & 2.57& 12.2(24)(06)\\
$D^+\to \pi^+\pi^+\pi^-\eta$&($-$40,\,38)&$514.6 \pm  25.7$ & 9.67& 34.1(17)(10)\\
$D^+\to \pi^+\pi^0\pi^0\eta$&($-$70,\,49)&$192.5 \pm  17.1$ & 3.86& 32.0(28)(17)\\
\hline\hline
\end{tabular}}
\end{table}

The systematic uncertainties arise from the sources discussed below and are estimated relative to the measured BFs. The uncertainties in the total ST yields come from the $M_{\rm BC}^{\rm tag}$ fits to the ST $\bar D$ candidates, which were determined as
0.5\% for both neutral and charged $\bar D$~\cite{epjc76,cpc40,bes3-pimuv}.
The systematic uncertainties of the tracking efficiencies are found to
be (0.2-0.5)\% per $K^\pm$ or $\pi^\pm$, while those for PID efficiencies are taken as (0.2-0.3)\% per $K^\pm$ or $\pi^\pm$, by using DT $D\bar D$ hadronic events.
The systematic uncertainty in $K_{S}^{0}$ reconstruction is estimated
to be 1.6\% per $K^0_S$ by using the
$J/\psi\to K^*(892)^{\mp}K^{\pm}$ and $J/\psi\to \phi K_S^{0}K^{\pm}\pi^{\mp}$ candidates~\cite{sysks}.
The systematic uncertainty of the $\pi^0$ reconstruction is assigned
as (0.7-0.8)\% per $\pi^0$ from studies of DT $D\bar D$ hadronic decay samples of $D^0\to K^-\pi^+$, $K^-\pi^+\pi^+\pi^-$ vs. $\bar D^0\to K^+\pi^-\pi^0$, $K^0_S\pi^0$~\cite{epjc76,cpc40}.
The systematic uncertainty for $\eta$ reconstruction is taken to be the
same as that for $\pi^0$.
The uncertainties of the quoted BFs of $K^0_S\to\pi^+\pi^-$, $\eta\to \gamma\gamma$, and $\pi^0\to \gamma\gamma$ decays are 0.07\%, 0.5\%, and 0.03\%~\cite{pdg2018}, respectively.

To estimate the systematic uncertainty in 2D fit, we repeat the fits by
varying the signal shape, the endpoint of the ARGUS function ($\pm 0.2$~MeV/$c^2$),
the fixed PBKG yield ($\pm 1\sigma$ of the quoted BF).
The systematic uncertainty of the $D\bar D$ opening angle requirement is assigned as 0.4\% by using the DT events of $D^0\to K^-\pi^+\pi^0$.
The systematic uncertainty due to the $\Delta E_{\rm sig}$ requirement is assigned to be 0.3\%,
which is the largest efficiency difference with and without smearing the data-MC Gaussian resolution of $\Delta E_{\rm sig}$ for signal MC events.
Here, the parameters of the Gaussian are obtained by using the DT samples of $D^0\to K^0_S\pi^0$, $K^-\pi^+\pi^0$, $K^-\pi^+\pi^0\pi^0$, and $D^+\to K^-\pi^+\pi^+\pi^0$.
The systematic uncertainties due to the choice of the $K^0_S$ sideband
and the $K^0_S/\omega/\eta^{(\prime)}/\phi$ rejection windows
are assigned by examining the changes of the BFs when varying the nominal $K^0_S$ sideband and rejection windows by $\pm5$ MeV/$c^2$.
The uncertainties due to limited MC statistics (0.3-1.1)\% are considered as a source of systematic uncertainty.

The systematic uncertainties in MC modeling are categorized into three cases.
For the $D\to \bar K\pi\eta$ decays, the differences between the DT efficiencies obtained with the
BODY3 and PHSP generators are assigned as the uncertainties.
For the decays whose efficiencies are estimated with PHSP generator,
the uncertainties are assigned by referring to
the largest change of the efficiencies among $D^0\to K^-\pi^+\eta$, $K^0_S\pi^0\eta$, and $D^+\to K^0_S\pi^+\eta$.
For the decays whose efficiencies are estimated with the mixed signal MC events, the systematic uncertainties are assigned as the change of the DT efficiency after removing the smallest component.

The $D^0\bar D^0$ pairs are produced coherently at the $\psi(3770)$.
For the decays $D^0\to K^0_S\pi^0\eta$, $K^+K^-\eta$, $K^0_SK^0_S\eta$, $K^0_S\pi^+\pi^-\eta$,  $K^0_S\pi^0\pi^0\eta$, and $\pi^+\pi^-\pi^0\eta$,  the measured BFs are affected by various $CP$ components due to quantum-correlation (QC) effects. The fractions of $CP+$ components in these decays
are examined by the $CP+$ tag of $D^0\to K^+K^-$ and $CP-$ tag of $D^0\to K^0_S\pi^0$, with the same method described in Ref.~\cite{QC-factor},
and the necessary parameters are taken from Refs.~\cite{R-ref1,R-ref2,R-ref3}.
The obtained impact of QC effects on the BFs ($f_{\rm QC}$) is shown in Table 1 of the supplemental material~\cite{supplemental}.
The signal efficiencies are corrected by the corresponding $f_{\rm QC}$ factors, the residual
statistical errors of $f_{\rm QC}$ are assigned as the systematic uncertainties.

For each signal decay, the total systematic uncertainty is obtained by
adding the above effects in quadrature.
The systematic uncertainties for the various signal decays are given in Table 2 of the supplemental material~\cite{supplemental} and the individual absolute systematic errors are summarized in Table~\ref{tab:DT} of text.

For the six decay modes with the highest yields, the BFs of $D$ and $\bar D$ decays,
${\mathcal B}^+_{\rm sig}$ and ${\mathcal B}^-_{\overline{\rm sig}}$, are measured separately.
Their asymmetry is determined by
${{\mathcal A}_{CP}^{\rm sig}}=\frac{{\mathcal B}^+_{\rm sig}-{\mathcal B}^-_{\overline{\rm sig}}}{{\mathcal B}^+_{\rm sig}+{\mathcal B}^-_{\overline{\rm sig}}}$.
The obtained BFs and asymmetries are summarized in Table~\ref{tab:CP}.
We find no statistically significant $CP$ violation.
Several systematic uncertainties cancel in the asymmetry: the tracking and PID of $\pi^+\pi^-/K^+K^-$ pair, $K^0_S$ reconstruction, $\pi^0/\eta$ reconstruction, quoted BFs, $K^0_S$ sideband choice, $K^0_S/\omega/\eta^{(\prime)}/\phi$ rejection windows, MC modeling,
and strong phase of $D^0$ decays.
The other systematic uncertainties are estimated separately as above.

\begin{table}[htp]
\centering
\caption{\small
Charge-conjugated BFs (${\mathcal B}^+_{\rm sig}$ and ${\mathcal B}^-_{\overline{\rm sig}}$),
and their asymmetries (${{\mathcal A}_{CP}^{\rm sig}}$).
The first and second uncertainties are statistical and systematic, respectively, for ${ {\mathcal A}_{CP}^{\rm sig}}$; while uncertainties for ${\mathcal B}^+_{\rm sig}$ and ${\mathcal B}^-_{\overline{\rm sig}}$ are only statistical.
}\label{tab:CP}
\resizebox{!}{1.7cm}{
\begin{tabular}{lrrc}
  \hline\hline
\multicolumn{1}{c} {Decay} & \multicolumn{1}{c} {${\mathcal B}^+_{\rm sig}$($\times 10^{-4}$)}  & \multicolumn{1}{c} {${\mathcal B}^-_{\overline{\rm sig}}$($\times 10^{-4}$)}  & ${ {\mathcal A}_{CP}^{\rm sig}}$ (\%) \\  \hline
$D^0\to K^-\pi^+\eta$       &$182.1\pm3.5$&$189.1\pm3.6$&$-1.9\pm1.3\pm1.0$\\
$D^0\to K^0_S\pi^0\eta$     &$ 98.4\pm4.8$&$106.3\pm5.1$&$-3.9\pm3.2\pm0.8$\\
$D^0\to K^-\pi^+\pi^0\eta$  &$ 41.7\pm2.7$&$ 48.8\pm3.2$&$-7.9\pm4.8\pm2.5$\\
$D^0\to \pi^+\pi^-\pi^0\eta$&$ 29.8\pm2.2$&$ 33.3\pm2.5$&$-5.5\pm5.2\pm2.4$\\
$D^+\to K^0_S\pi^+\eta$     &$129.9\pm5.3$&$132.3\pm5.4$&$-0.9\pm2.9\pm1.0$\\
$D^+\to \pi^+\pi^+\pi^-\eta$&$ 35.4\pm2.4$&$33.7\pm2.4$&$+2.5\pm5.0\pm1.6$\\
\hline\hline
\end{tabular}
}
\end{table}

In summary, with $2.93\,\rm fb^{-1}$ of data taken at $\sqrt{s}=3.773$\,GeV with the BESIII detector, we report the first measurements of the absolute BFs of fourteen exclusive $D^{0(+)}$ decays to $\eta$.
Summing over the BFs measured in this work, and using the world averaged values of other known decays~\cite{pdg2018}, the total BFs of all the exclusive
$D^0$ and $D^+$ decays to $\eta$ are determined to be
$(8.62\pm0.35)\%$ and $(4.68\pm0.18)\%$, respectively. Here, the
systematic uncertainties of $ N_{\rm ST}$, $K^\pm/\pi^\pm$ tracking
and PID, $K^0_S$ and $\eta$ reconstruction, and the quoted BFs are correlated. They are consistent with the corresponding inclusive rates $(9.5\pm0.9)\%$ and $(6.5\pm0.7)\%$
within $0.9\sigma$ and $2.5\sigma$, respectively, leaving little room for other exclusive decays involving $\eta$.
The reported BFs provide key inputs for accurate background estimations in LFU tests with semileptonic $B$ decays, which are crucial to explore possible new physics beyond the SM.
The obtained ${\mathcal B}(D^0\to K^-\pi^+\eta)$ agrees with the recent Belle result~\cite{belle-kpieta,pdg2018} within $1.3\sigma$, with precision improved twofold.
Our ${\mathcal B}(D^0\to K^0_S\pi^0\eta)$ is greater than CLEO's result~\cite{cleo-kspi0eta,pdg2018} by $3.7\sigma$.
Combining the measured ${\mathcal B}(D^0\to K^0_S\pi^0\eta)$ with the fit fraction ${\mathcal B}(D^0\to \bar K^*(892)^0\eta,\bar K^*(892)^0\to K^0_S\pi^0)/{\mathcal B}(D^0\to K^0_S\pi^0\eta)$ from CLEO~\cite{cleo-kspi0eta}, we find ${\mathcal B}(D^0\to \bar K^*(892)^0\eta)={(1.77\pm0.44)}\%$, where the uncertainty is dominated by the fit fraction.
This deviates from various theoretical calculations~\cite{ref5,theory_1,theory_2} by 1.9-2.9$\sigma$.
Future amplitude analyses of these decays at BESIII~\cite{bes3-white-paper} and Belle II~\cite{belle2-white-paper} will open a window to extract more two-body hadronic $D$ decays,
which are important to understand quark U-spin and SU(3)-flavor
symmetry breaking effects, and will be beneficial for the predictions of $D^0\bar D^0$ mixing
and $CP$ violation in $D$ decays~\cite{ref5,theory_1,theory_2}.
In addition, we determine the asymmetries of the charge-conjugated BFs for the six $D$ decays with highest yields,
and we find no statistically significant $CP$ violation.

The BESIII collaboration thanks the staff of BEPCII and the IHEP computing center for their strong support. This work is supported in part by National Key Basic Research Program of China under Contract No. 2015CB856700; National Natural Science Foundation of China (NSFC) under Contracts Nos.~11775230, 11475123, 11625523, 11635010, 11735014, 11822506, 11835012, 11935015, 11935016, 11935018, 11961141012; the Chinese Academy of Sciences (CAS) Large-Scale Scientific Facility Program; Joint Large-Scale Scientific Facility Funds of the NSFC and CAS under Contracts Nos.~U1532101, U1932102, U1732263, U1832207; CAS Key Research Program of Frontier Sciences under Contracts Nos. QYZDJ-SSW-SLH003, QYZDJ-SSW-SLH040; 100 Talents Program of CAS; INPAC and Shanghai Key Laboratory for Particle Physics and Cosmology; ERC under Contract No. 758462; German Research Foundation DFG under Contracts Nos. Collaborative Research Center CRC 1044, FOR 2359; Istituto Nazionale di Fisica Nucleare, Italy; Ministry of Development of Turkey under Contract No. DPT2006K-120470; National Science and Technology fund; STFC (United Kingdom); The Knut and Alice Wallenberg Foundation (Sweden) under Contract No. 2016.0157; The Royal Society, UK under Contracts Nos. DH140054, DH160214; The Swedish Research Council; U. S. Department of Energy under Contracts Nos. DE-FG02-05ER41374, DE-SC-0012069.


\clearpage
\appendix
\onecolumngrid
\section*{Supplemental material}
\setcounter{table}{0}
\setcounter{figure}{0}

\begin{figure*}[htp]
  \centering
\includegraphics[width=0.4\linewidth]{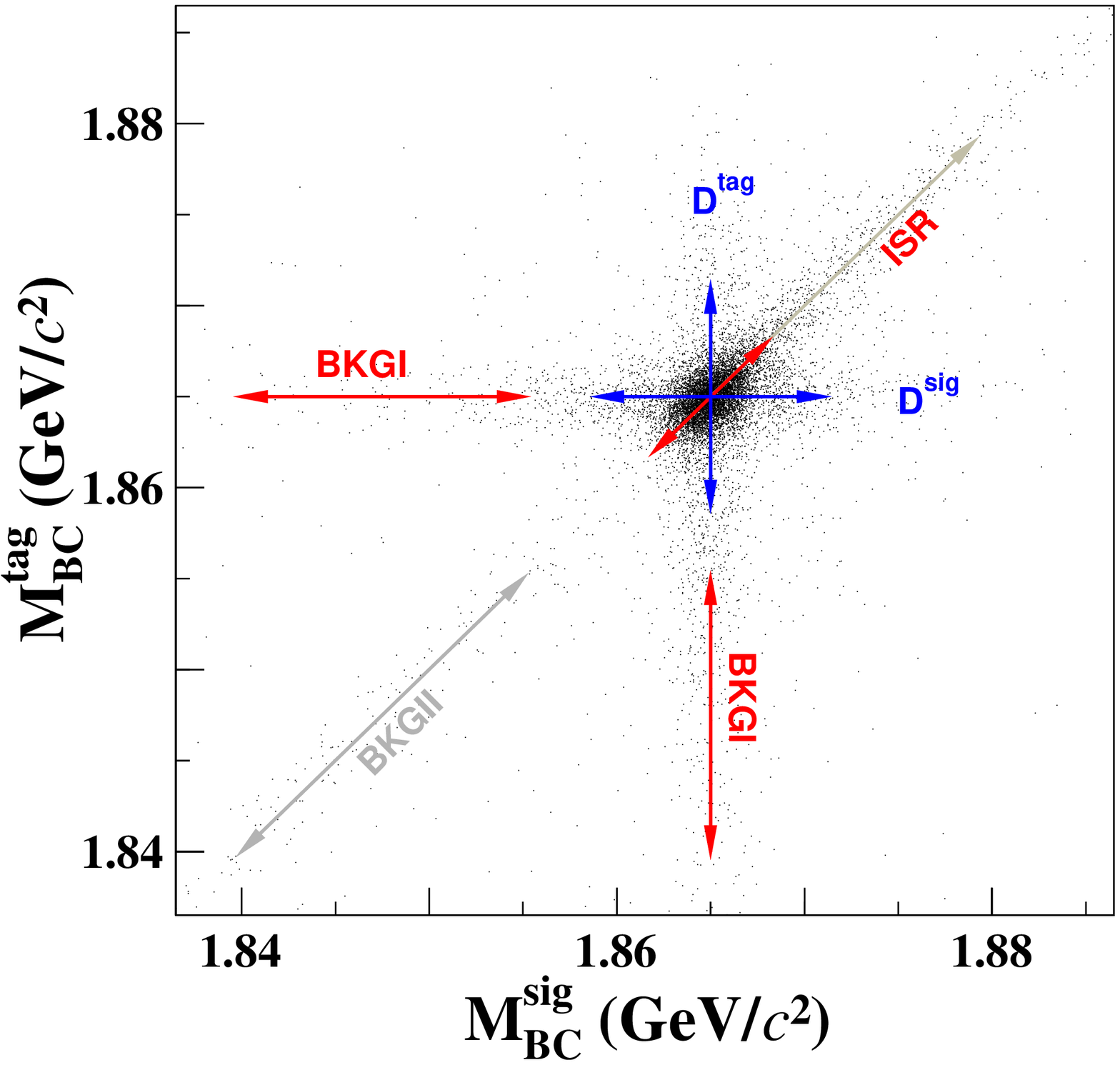}
  \caption{
The $M^{\rm tag}_{\rm BC}$ vs. $M^{\rm sig}_{\rm BC}$ distribution of the accepted DT candidate events.}
\label{fig:2D}
\end{figure*}

Figure \ref{fig:2D} shows illustration of the $M^{\rm tag}_{\rm BC}$ vs. $M^{\rm sig}_{\rm BC}$ distribution of the accepted DT candidate events.

\begin{figure*}[htp]
  \centering
\includegraphics[width=0.8\linewidth]{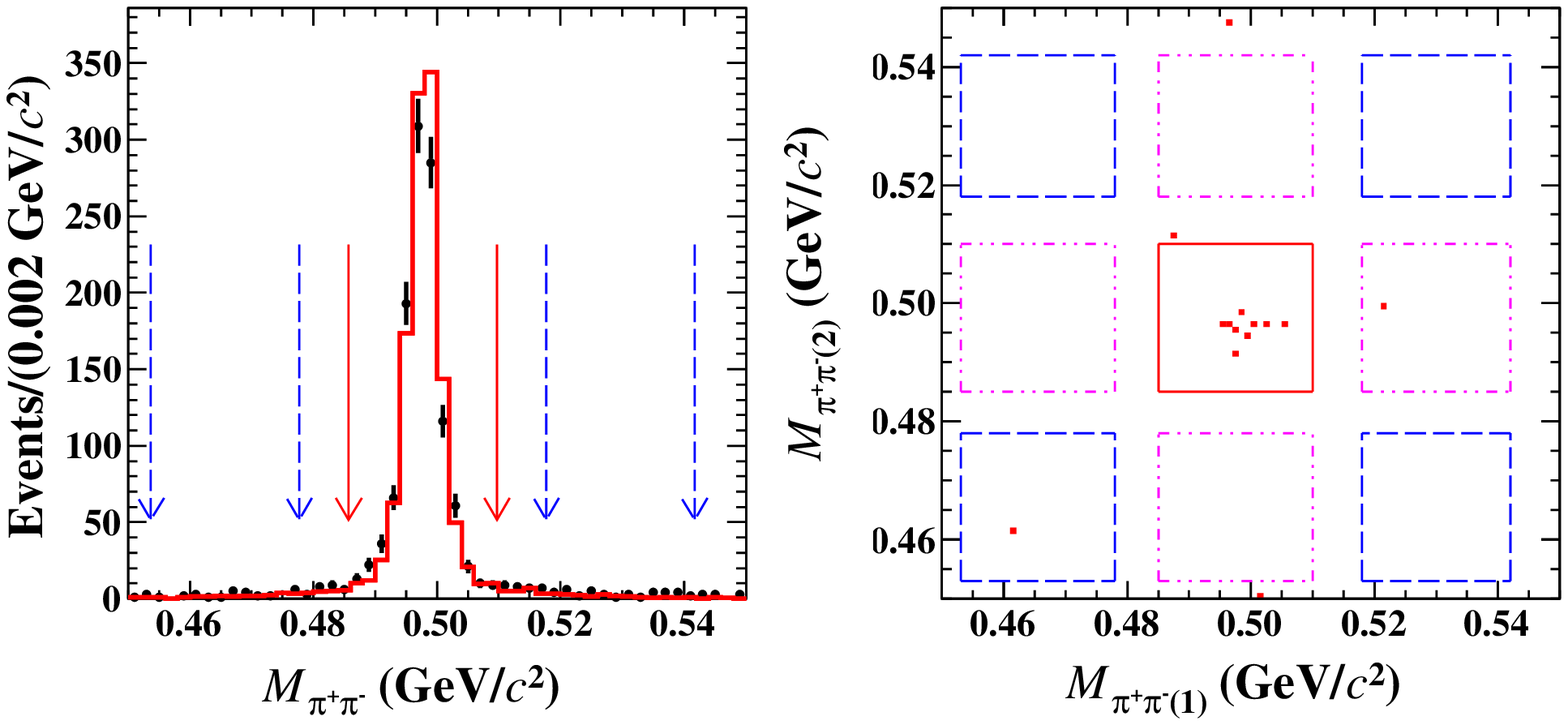}
  \put(-140,160){\bf (b)}
  \put(-350,160){\bf (a)}
  \caption{
(a)~Comparison of the $M_{\pi^+\pi^-}$ distributions of the $D^0\to K^0_S\pi^0\eta$ candidate events
between data (dots with error bars) and the inclusive MC sample (histogram).
Pairs of the solid~(dashed) arrows denote the 1D $K^0_S$ signal~(sideband) regions.
(b)~Distribution of $M_{\pi^+\pi^-(1)}$ vs. $M_{\pi^+\pi^-(2)}$ for the $D^0\to K^0_SK^0_S\eta$ candidate events in data.
Solid box denotes the 2D signal region.
Dot-dashed~(dashed) boxes indicate the 2D sideband 1~(2) regions.
}
\label{fig:mks}
\end{figure*}

Figure~\ref{fig:mks} shows the definitions of 1D and 2D $K^0_S$ signal and sideband regions.

Table~\ref{tab:QC} summarizes
the ST yields of $CP\pm$ tags from the fits to the $M^{\rm tag}_{\rm BC}$ distributions of the accepted ST candidates,
the DT yields tagged by $CP\pm$ tags from the 2D fits to the $M^{\rm tag}_{\rm BC}$ vs. $M^{\rm sig}_{\rm BC}$ distributions of the accepted DT candidates, and  the QC factors obtained with the
same method as described in Ref.~\cite{QC-factor} and the necessary parameters quoted from Refs.~\cite{R-ref1,R-ref2,R-ref3}.
No DT events are observed from the $D^0\to K^+K^-\eta$ and $K^0_SK^0_S\eta$ decays.
The systematic uncertainties arising from QC effects are directly assigned as the averaged strong-phase factor $C_f$ by the flavor tag yields.

Table~\ref{tab:sys} summarizes the systematic uncertainties for various sources in the measurements of BFs, which
are assigned relative to the measured BFs.
For each signal decay, the total uncertainty is obtained by quadratically adding all errors.

\begin{table*}[htp]
\centering
\caption{
Summary of the ST yields of $CP\pm$ tags ($S^\pm_{\rm measured}$),
the DT yields tagged by $CP\pm$ tags ($M^\mp_{\rm measured}$), and the QC factor ($f_{\rm QC}$).
The errors are statistical only.}
\label{tab:QC}
\centering
\begin{tabular}{c|cc|c|c}
  \hline
\multirow{2}{*}{$CP$ tag}& $S^+_{\rm measured}$&$S^-_{\rm measured}$&\multirow{2}{*}{}&\multirow{2}{*}{}\\
\cline{2-3}
&57779$\pm$287&70512$\pm$311&&\\ \hline
\multicolumn{1}{c|}{Decay} &$M^-_{\rm measured}$&$M^+_{\rm measured}$& $f_{\rm QC}$ & Uncertainty (\%)\\ \hline
$D^0\to K^0_S\pi^0\eta$     &$2.4^{+1.6}_{-2.0}$     &$67.6\pm8.3$       &$0.942^{+0.007}_{-0.008}$&0.8\\
$D^0\to K^+K^-\eta$         &$0$                     &$0$                &--&7.4\\
$D^0\to K^0_SK^0_S\eta$     &$0$                     &$0$                &--&7.4\\
$D^0\to K^0_S\pi^+\pi^-\eta$&$19.8\pm4.7$            &$2.0^{+0.9}_{-1.1}$&$1.057^{+0.013}_{-0.013}$&1.3\\
$D^0\to K^0_S\pi^0\pi^0\eta$&$5.4^{+2.8}_{-2.4}$     &$0$                &$1.073^{+0.065}_{-0.040}$&6.5\\
$D^0\to \pi^+\pi^-\pi^0\eta$&$13.6\pm4.8$            &$18.8\pm4.4$       &$0.993^{+0.008}_{-0.008}$&0.8\\
\hline
\end{tabular}
\end{table*}

\begin{table*}[htp]
\centering
\caption{
Systematic uncertainties (\%) in the measurements of the BFs.}
\label{tab:sys}
\centering
\resizebox{\textwidth}{30mm}{
\begin{tabular}{c|cccccccc|cccccc}
  \hline
\multirow{2}{*}{Sources}&\multicolumn{8}{c|}{$D^0\to$}&\multicolumn{6}{c}{$D^+\to$}\\
&
$K^-\pi^+\eta$&$K^0_S\pi^0\eta$&$K^+K^-\eta$&$K^0_SK^0_S\eta$&$K^-\pi^+\pi^0\eta$&$K^0_S\pi^+\pi^-\eta$&$K^0_S\pi^0\pi^0\eta$&$\pi^+\pi^-\pi^0\eta$&$K^0_S\pi^+\eta$&$K^0_SK^+\eta$& $\pi^+\pi^+\pi^-\eta$&$K^0_S\pi^+\pi^0\eta$&$K^-\pi^+\pi^+\eta$&$\pi^+\pi^0\pi^0\eta$\\
  \hline
$N^{\rm tot}_{\rm ST}$                         &0.5 &0.5 &0.5 &0.5 &0.5 &0.5 &0.5 &0.5 &0.5 &0.5 &0.5 &0.5 &0.5 &0.5\\
$K^\pm/\pi^\pm$ tracking                       &0.7 &--  &1.0 &--  &0.9 &0.7 &--  &0.6 &0.2 &0.5 &1.2 &0.4 &0.9 &0.3\\
$K^\pm/\pi^\pm$ PID                            &0.4 &--  &0.5 &--  &0.4 &0.4 &--  &0.4 &0.2 &0.3 &0.7 &0.2 &0.6 &0.2\\
$\pi^0/\eta$ reconstruction                    &0.7 &1.5 &0.8 &0.8 &1.4 &0.8 &2.2 &1.5 &0.7 &0.8 &0.7 &1.5 &0.7 &2.2\\
$K^0_S$ reconstruction                         &--  &1.6 &--  &3.2 &--  &1.6 &1.6 &--  &1.6 &1.6 &--  &1.6 &--  &-- \\
Quoted $\mathcal B$                            &0.5 &0.5 &0.5 &0.5 &0.5 &0.5 &0.5 &0.5 &0.5 &0.5 &0.5 &0.5 &0.5 &0.5\\
2D fit                                         &0.5 &1.0 &5.3 &10.5&2.4 &0.7 &2.9 &2.6 &0.9 &3.6 &1.9 &4.5 &2.1 &3.2\\
$D\bar D$ angle                                &0.4 &0.4 &0.4 &0.4 &0.4 &0.4 &0.4 &0.4 &0.4 &0.4 &0.4 &0.4 &0.4 &0.4\\
$\Delta E^{\rm sig}$ requirement               &0.3 &0.3 &0.3 &0.3 &0.3 &0.3 &0.3 &0.3 &0.3 &0.3 &0.3 &0.3 &0.3 &0.3\\
$K_{S}^{0}/\eta/\omega/\eta^{\prime}$ rejection&--  &--  &--  &--  &--  &3.7 &5.9 &2.3 &--  &--  &--  &--  &1.9 &2.9\\
$K_{S}^{0}$ sideband                           &--  &0.5 &--  &--  &--  &0.6 &--  &--  &0.2 &--  &--  &--  &--  &-- \\
MC statistics                                  &0.3 &0.5 &0.4 &0.6 &0.5 &0.6 &1.1 &0.5 &0.4 &0.5 &0.4 &0.7 &0.4 &0.7\\
MC generator                                   &0.5 &0.8 &0.9 &0.9 &1.2 &1.5 &0.9 &1.8 &0.9 &0.9 &1.5 &0.9 &1.2 &1.9\\
Strong phase of neutral $D$                    &--  &0.8 &7.4 &7.4 &--  &1.3 &6.5 &0.8 &--  &--  &--  &--  &--  &-- \\
\hline
Total                                          &1.6 &2.9 &9.3&13.3 &3.4 &3.8 &8.3 &4.4 &2.4 &4.2 &3.0 &5.2 &3.4 &5.3\\
\hline
\end{tabular}}
\end{table*}


\begin{thebibliography}{**}

\bibitem{babar_1}
J. P. Lees {\it et al.} (BaBar Collaboration),
\href{https://journals.aps.org/prl/abstract/10.1103/PhysRevLett.109.101802}{Phys. Rev. Lett. {\bf 109}, 101802 (2012).}

\bibitem{babar_2}
J. P. Lees {\it et al.} (BaBar Collaboration),
\href{https://journals.aps.org/prd/abstract/10.1103/PhysRevD.88.072012}{Phys. Rev. D {\bf 88}, 072012 (2013).}

\bibitem{lhcb_1}
R. Aaij {\it et al.} (LHCb Collaboration),
\href{https://journals.aps.org/prl/abstract/10.1103/PhysRevLett.115.159901}{Phys. Rev. Lett. {\bf 115}, 111803 (2015).}

\bibitem{belle2015}
M. Huschle {\it et al.} (Belle Collaboration),
\href{https://journals.aps.org/prd/abstract/10.1103/PhysRevD.92.072014}{Phys. Rev. D {\bf 92}, 072014 (2015).}

\bibitem{belle2016}
Y. Sato {\it et al.} (Belle Collaboration),
\href{https://journals.aps.org/prd/abstract/10.1103/PhysRevD.94.072007}{Phys. Rev. D {\bf 94}, 072007 (2016).}

\bibitem{lhcb_1a}
R. Aaij {\it et al.} (LHCb Collaboration),
\href{https://journals.aps.org/prd/abstract/10.1103/PhysRevD.97.072013}{Phys. Rev. D {\bf 97}, 072013 (2018).}

\bibitem{belle2019}
A. Abdesselam {\it et al.} (Belle Collaboration),
\href{https://arxiv.org/abs/1904.08794}{arXiv:1904.08794.}

\bibitem{belle2019_2} G. Garia {\it et al.} (Belle collaboration),
\href{https://journals.aps.org/prl/abstract/10.1103/PhysRevLett.124.161803}{Phys. Rev. Lett. {\bf 124}, 161803 (2020).}

\bibitem{hflav2018}
Y. Amhis {\it et al.} (Heavy Flavor Averaging Group),
\href{https://arxiv.org/abs/1909.12524}{arXiv:1909.12524.}

\bibitem{lhcbnote}
R. Aaij {\it et al.} (LHCb Collaboration),
\href{http://cds.cern.ch/record/2223391?ln=zh_CN}{Synergy of BESIII and LHCb physics programmes, LHCb-PUB-2016-025.}

\bibitem{pdg2018}
M. Tanabashi {\it et al.} (Particle Data Group),
\href{https://journals.aps.org/prd/abstract/10.1103/PhysRevD.98.030001}{Phys. Rev. D {\bf 98}, 030001 (2018) and 2019 update.}

\bibitem{belle-kpieta}
Y. Q. Chen {\it et al}. (Belle Collaboration),
\href{https://arxiv.org/abs/2003.07759}{arXiv:2003.07759.}

\bibitem{cleo-kspi0eta}
P. Rubin {\it et al}. (CLEO Collaboration),
\href{https://journals.aps.org/prl/abstract/10.1103/PhysRevLett.93.111801}{Phys. Rev. Lett. {\bf 93}, 111801 (2004).}

\bibitem{Kingsley}
R. L. Kingsley, S. B. Treiman, F. Wilczek and A. Zee,
\href{https://journals.aps.org/prd/abstract/10.1103/PhysRevD.11.1919}{Phys. Rev. D {\bf 11}, 1919 (1975).}

\bibitem{Gronau}
M. Gronau,
\href{https://doi.org/10.1016/j.physletb.2014.01.035}{Phys. Lett. B {\bf 730}, 221 (2014);}
\href{https://doi.org/10.1016/j.physletb.2014.06.055}{{\bf 735}, 282(E) (2014).}

\bibitem{zzxing}
Z. Z. Xing,
\href{https://doi.org/10.1142/S0217732319502389}{Mod. Phys. Lett. A {\bf 34}, 1950238 (2019).}

\bibitem{theory_a}
H. J. Lipkin,
\href{https://journals.aps.org/prl/abstract/10.1103/PhysRevLett.46.1307}{Phys. Rev. Lett. {\bf 46}, 1307 (1981).}

\bibitem{theory_1}
H. Y. Cheng and C. W. Chiang,
\href{https://journals.aps.org/prd/abstract/10.1103/PhysRevD.81.074021}{ Phys. Rev. D {\bf 81}, 074021 (2010).}

\bibitem{theory_2}
Q. Qin, H. N. Li, C. D. L\"u, and F. S. Yu,
\href{https://journals.aps.org/prd/abstract/10.1103/PhysRevD.89.054006}{Phys. Rev. D {\bf 89}, 054006 (2014).}

\bibitem{theory_3}
H. Y. Cheng, C. W. Chiang, and A. L. Kuo,
\href{https://journals.aps.org/prd/abstract/10.1103/PhysRevD.93.114010}{Phys. Rev. D {\bf 93}, 114010 (2016).}

\bibitem{theory_4}
W. Kwong and S. P. Rosen,
\href{https://doi.org/10.1016/0370-2693(93)91843-C}{Phys. Lett. B {\bf 298}, 413 (1993).}

\bibitem{theory_5}
Y. Grossman and D. J. Robinson,
\href{https://doi.org/10.1007/JHEP04(2013)067}{J. High Energy Phys. {\bf 1304} (2013) 067.}

\bibitem{ref5}
H. N. Li, C. D. L\"u, and F. S. Yu,
\href{https://journals.aps.org/prd/abstract/10.1103/PhysRevD.86.036012}{Phys. Rev. D {\bf 86}, 036012 (2012).}


\bibitem{ref1}
I. I. Bigi, A. Paul, and S. Recksiegel,
\href{https://doi.org/10.1007/JHEP06(2011)089}{J. High Energy Phys. {\bf 1106} (2011) 089.}

\bibitem{ref2}
G. Isidori, J. F. Kamenik, Z. Ligeti, and G. Perez,
\href{https://doi.org/10.1016/j.physletb.2012.03.046}{Phys. Lett. B {\bf 711}, 46 (2011).}

\bibitem{ref3}
J. Brod, A. L. Kagan, and J. Zupan,
\href{https://journals.aps.org/prd/abstract/10.1103/PhysRevD.86.014023}{Phys. Rev. D {\bf 86}, 014023 (2012).}

\bibitem{ref4}
H. Y. Cheng and C. W. Chiang,
\href{https://journals.aps.org/prd/abstract/10.1103/PhysRevD.86.014014}{Phys. Rev. D {\bf 86}, 014014 (2012).}

\bibitem{ref6}
H. Y. Cheng and C. W. Chiang,
\href{https://journals.aps.org/prd/abstract/10.1103/PhysRevD.100.093002}{Phys. Rev. D {\bf 100}, 093002 (2019).}

\bibitem{ref7}
H. N. Li, C. D. L\"u, and F. S. Yu,
\href{https://arxiv.org/abs/1903.10638}{arXiv:1903.10638.}

\bibitem{lhcb_D_CP}
R. Aaij {\it et al}. (LHCb Collaboration),
\href{https://journals.aps.org/prl/abstract/10.1103/PhysRevLett.122.211803}{Phys. Rev. Lett {\bf 122}, 211803 (2019).}

\bibitem{lum_bes3}
M. Ablikim {\it et al.} (BESIII Collaboration),
\href{https://iopscience.iop.org/article/10.1088/1674-1137/37/12/123001}{Chin. Phys. C {\bf 37}, 123001 (2013);}
\href{https://doi.org/10.1016/j.physletb.2015.11.043}{Phys. Lett. B {\bf 753}, 629 (2016).}

\bibitem{BESIII}
M. Ablikim {\it et al.} (BESIII Collaboration),
\href{https://doi.org/10.1016/j.nima.2009.12.050}{Nucl. Instrum. Meth. A {\bf 614}, 345 (2010).}

\bibitem{geant4}
S. Agostinelli {\it et al.} (GEANT4 Collaboration),
\href{https://doi.org/10.1016/S0168-9002(03)01368-8}{Nucl. Instrum. Meth. A {\bf 506}, 250 (2003).}

\bibitem{kkmc}
S. Jadach, B. F. L. Ward, and Z. Was,
\href{https://journals.aps.org/prd/pdf/10.1103/PhysRevD.63.113009} {Phys. Rev. D {\bf 63}, 113009 (2001);}
\href{https://doi.org/10.1016/S0010-4655(00)00048-5}{Comput. Phys. Commun.  {\bf 130}, 260 (2000).}

\bibitem{evtgen}
D.~J.~Lange,
\href{https://doi.org/10.1016/S0168-9002(01)00089-4} {Nucl. Instrum. Meth. A {\bf 462}, 152 (2001);}
R.~G.~Ping,
\href{https://doi.org/10.1088/1674-1137/32/8/001}{Chin. Phys. C {\bf 32}, 599 (2008).}

\bibitem{lundcharm}
J. C. Chen, G. S. Huang, X. R. Qi, D. H. Zhang, and Y. S. Zhu,
\href{https://journals.aps.org/prd/abstract/10.1103/PhysRevD.62.034003}{Phys. Rev. D {\bf 62}, 034003 (2000).}

\bibitem{epjc76}
M. Ablikim {\it et al.} (BESIII Collaboration),
\href{https://doi.org/10.1140/epjc/s10052-016-4198-2} {Eur. Phys. J. C {\bf 76}, 369 (2016).}

\bibitem{cpc40}
M. Ablikim {\it et al.} (BESIII Collaboration),
\href{https://iopscience.iop.org/article/10.1088/1674-1137/40/11/113001}{Chin. Phys. C {\bf 40}, 113001 (2016).}

\bibitem{bes3-pimuv}
M. Ablikim {\it et al.} (BESIII Collaboration),
\href{https://journals.aps.org/prl/abstract/10.1103/PhysRevLett.121.171803}{Phys. Rev. Lett. {\bf 121}, 171803 (2018).}

\bibitem{bes3-Dp-K1ev}
M. Ablikim {\it et al.} (BESIII Collaboration),
\href{https://journals.aps.org/prl/abstract/10.1103/PhysRevLett.123.231801}{Phys. Rev. Lett. {\bf 123}, 231801 (2019).}

\bibitem{bes3-etaetapi}
M. Ablikim {\it et al.} (BESIII Collaboration),
\href{https://journals.aps.org/prd/pdf/10.1103/PhysRevD.101.052009}{Phys. Rev. D {\bf 101}, 052009 (2020).}

\bibitem{bes3-omegamuv}
M. Ablikim {\it et al.} (BESIII Collaboration),
\href{https://journals.aps.org/prd/pdf/10.1103/PhysRevD.101.072005}{Phys. Rev. D {\bf 101}, 072005 (2020).}

\bibitem{bes3-etamuv}
M. Ablikim {\it et al.} (BESIII Collaboration),
\href{https://journals.aps.org/prl/abstract/10.1103/PhysRevLett.124.231801}{Phys. Rev. Lett. {\bf 124}, 231801 (2020).}

\bibitem{supplemental}
See Supplemental Material for
an illustration of the $M_{\rm BC}^{\rm tag}$ vs.~$M_{\rm BC}^{\rm sig}$ distribution,
the definitions of 1D and 2D signal and sideband regions of $K^0_S$,
the correction factors due to QC effects, and
the systematic uncertainties of various sources in the BF measurements.

\bibitem{ARGUS}
H. Albrecht {\it et al.} (ARGUS Collaboration),
\href{https://doi.org/10.1016/0370-2693(90)91293-K}{Phys. Lett. B {\bf 241}, 278 (1990).}

\bibitem{Stenson}
K. Stenson, \href{https://arxiv.org/abs/physics/0605236}{arXiv:physics/0605236.}

\bibitem{sysks}
M. Ablikim {\it et al}. (BESIII Collaboration),
\href{https://journals.aps.org/prd/abstract/10.1103/PhysRevD.92.112008}{Phys. Rev. D {\bf 92}, 112008 (2015).}

\bibitem{QC-factor}
M. Ablikim {\it et al.} (BESIII Collaboration),
\href{https://journals.aps.org/prd/abstract/10.1103/PhysRevD.100.072006}{Phys. Rev. D {\bf 100}, 072006 (2019).}

\bibitem{R-ref1}
T.Gershon, J. Libby, and G. Wilkinson,
\href{https://doi.org/10.1016/j.physletb.2015.08.063} {Phys. Lett. B {\bf 750}, 338 (2015).}

\bibitem{R-ref2}
T. Evans {\it et al.},
\href{https://doi.org/10.1016/j.physletb.2016.04.037}{Phys. Lett. B {\bf 757}, 520 (2016);}
\href{https://doi.org/10.1016/j.physletb.2016.11.021}{{\bf 765}, 402(E) (2017).}

\bibitem{R-ref3}
Heavy Flavor Averaging Group (HFLAV),
\href{http://www.slac.stanford.edu/xorg/hflav/charm/}{(http://www.slac.stanford.edu/xorg/hflav/charm/).}

\bibitem{bes3-white-paper} M. Ablikim {\it et al.} (BESIII Collaboration),
\href{http://cpc.ihep.ac.cn/article/doi/10.1088/1674-1137/44/4/040001} {Chin. Phys. C {\bf 44}, 040001 (2020).}

\bibitem{belle2-white-paper}
E. Kou {\it et al.} (Belle II Collaboration),
\href{https://doi.org/10.1093/ptep/ptz106} {PTEP {\bf 2019}, 123C01 (2019).}

\end{thebibliography}
\end{document}